# Limiting Behavior of LQ Deterministic Infinite Horizon Nash Games with Symmetric Players as the Number of Players goes to Infinity

G.P.Papavassilopoulos
Dept. of Electrical and Computer Engineering
National Technical University of Athens
9 Iroon Polytechniou Str., 157 73
Athens, Greece
(yorgos@netmode.ntua.gr)

ABSTRACT
A Linear Quadratic Deterministic Continuous Time Game with many symmetric players is considered and the Linear Feedback Nash strategies are studied as the number of players goes to infinity. We show that under some conditions the limit of the solutions exists and can be used to approximate the case with a finite but large number of players. It is shown that in the limit each player acts as if he were faced with one player only, who represents the average behavior of the others.

## 1. INTRODUCTION

In most papers in game theory there may be few or many players and solutions and properties are usually studied for a fixed such number. There are also papers where the multitude of players is infinite see Refs.8, 9 as well as many recent papers on games played by an infinite number of automata each one of which interacts with a finite number of neighbors. In the recent literature of mean field games (Refs. 10-16), stochastic games are considered with an infinite number of players who are modeled in a statistical sense. There are also papers where a version of the problem is solved for a finite number of players and the validity or evolution of some property is studied as the number of players goes to infinity (Refs. 5-7).The present paper belongs to this last category. We consider a finite number of players involved in a Linear Quadratic, deterministic, continuous time, infinite horizon Nash Game. To each one corresponds his control and a part of the overall state called his state. The game is symmetric in the sense that the evolution of the part of the state that pertains to each player is influenced in a symmetric manner by the controls and states of the others. Finally the cost of each player depends only on his state and control and it is also symmetric for all the players. We employ the Linear Feedback Nash equilibrium concept and study how the solutions change as the number of players goes to infinity. It should be noticed that the present work is related to recent work on mean field games, Refs. 12-14, and in particular 15 and 16, but there are essential differences. First of all our problem is deterministic whereas those in Refs. 15 and 16 are stochastic. In Ref.15, Nash Open Loop finite time horizon solutions are studied for stochastic LQ Games, whereas we consider Linear Feedback Nash Solutions for the infinite time horizon that satisfy the Principle of Dynamic Programming. Also Ref.16 considers for a stochastic LQ game the Nash Linear Feedback solution like we do, but there the scalar case is considered only whereas we consider the deterministic set up for the matrix case. For the particular structure of game chosen, we show that the limit exists and can be used to approximate the case where the number of players is very large. The conditions for this to happen are related to the existence and properties of solutions of a generalized Ricatti equation, which yields stabilizing solutions for the overall game. The study of an associated not standard Hamiltonian problem is shown to be crucial for the whole analysis .The resulting infinite number of players case shows that under some conditions, each player essentially acts as if he were faced with one fictitious player who represents the average behavior of all the others or the market as a whole.

## 2. PROBLEM STATEMENT

Let us consider a dynamical system with state $x = (x_1, x_2, ... x_M)$. Each $x_i \in R^n$ evolves as

$$\frac{dx_i}{dt} = A_1 x_i + \frac{A_2}{M}(x_1 + x_2 + ... + x_M) + B_1 u_i + \frac{B_2}{M}(u_1 + u_2 + ... + u_M) \quad (1)$$

where $u_i \in R^m$ is the control of the i-th player. We have $M \geq 2$ players and as the state equation shows we can think of $x_i$ as the part of the state, or subsystem, which pertains to the player i.

The cost of the i-th player is given by:

$$J_i = \frac{1}{2} \int_0^\infty (x_i^T Q x_i + u_i^T u_i) dt \quad (2)$$

The matrices $A_1(n \times n), A_2(n \times n), B_1(n \times m), B_2(n \times m), Q(m \times m), Q = Q^T \geq 0$ are real and constant. All the $x_i$'s have the same dimension. Similarly all the $u_i$'s have the same dimension. Notice that all the subsystems have the same $A_1, A_2, B_1, B_2, Q = Q^T \geq 0$ and thus we have symmetry among them. The constants $\frac{A_2}{M}, \frac{B_2}{M}$ represent the explicit coupling between the subsystem of player i and the subsystems and controls of the other players. We are thus faced with a many player game problem for which the Nash equilibrium will be sought. The multitude M of the players will be considered as a given constant. Our main objective is to derive the solutions and study the resulting behavior in terms of both state and costs as M grows towards infinity.

The solution concept that we will employ is the Nash equilibrium where the players use linear strategies in the current state $x = (x_1, x_2, ... x_M)$ and Dynamic Programming holds (i.e. Linear Feedback Strategies in the parlance of the literature Refs 1-4).

**Definition 1** A set of linear strategies $u_i^* = L_i^* x, i = 1, 2 ..., M$, (where the $L_1^*, L_2^*, ..., L_M^*$ are constant $m \times n$ matrices), are said to be in Nash Equilibrium if for any initial condition $x(0) = (x_1(0), x_2(0), ... x_M(0))$ all the $J_i(u_1^*, u_2^*, ..., u_{i-1}^*, u_i^*, u_{i+1}^*, ..., u_M^*), i = 1, 2, ..., M$ are finite and

$J_i(u_1^*, u_2^*, ..., u_{i-1}^*, u_i^*, u_{i+1}^*, ..., u_M^*) \leq J_i(u_1^*, u_2^*, ..., u_{i-1}^*, u_i, u_{i+1}^* ..., u_M^*),$ for any other $u_i, i = 1, 2, ..., M$

(equivalently: for any other $L_i$).

The finiteness of the costs can be warranted by assuming that the closed loop system is asymptotically stable i.e. the closed loop matrix has all its eigenvalues in the open left-hand plane.

**Definition 2** A set of linear strategies $u_i^* = L_i^* x, i = 1, 2 ..., M$, (where the $L_1^*, L_2^*, ..., L_M^*$ are constant $m \times n$ matrices), are said to be in an $\varepsilon(M, x(0))$ -Nash Equilibrium if for any initial condition $x(0) = (x_1(0), x_2(0), ... x_M(0))$ all the $J_i(u_1^*, u_2^*, ..., u_{i-1}^*, u_i^*, u_{i+1}^*, ..., u_M^*), i = 1, 2, ..., M$ are finite and

$J_i(u_1^*, u_2^*, ..., u_{i-1}^*, u_i^*, u_{i+1}^*, ..., u_M^*) \leq J_i(u_1^*, u_2^*, ..., u_{i-1}^*, u_i, u_{i+1}^* ..., u_M^*) + \varepsilon(M, x(0)),$ for any other $u_i, i = 1, 2, ..., M$

(equivalently: for any other $L_i$).

Clearly the notion of $\varepsilon(M, x(0))$ -Nash Equilibrium is of interest if the $\varepsilon(M, x(0))$ is small and its magnitude can be qualified in terms of its arguments, see Comment 6.

The reasons for the choice of this type of Nash equilibrium are several. We know that in the corresponding deterministic discrete time framework, many and perhaps nonlinear strategies may exist; but if we introduce some nondegenerate noise in the state equation, then only the linear ones survive (Markov or Perfect equilibria, Refs 1,2). In the continuous time case, a similar phenomenon appears as regularization due to noise, of the system of the Hamilton-Jacobi-Bellman equations that characterize (as sufficient conditions) the Nash closed loop no memory solutions in the deterministic case.

### 3. NASH SOLUTION

The Nash solution for each player i is sought in the form (3), i.e. restrict ourselves to Symmetric Linear Strategies. (It is known that for the infinite time problem considered here,there might exist nonsymmetric strategies. The reason we consider symmetric ones is because our game is symmetric in the way the state equations and the costs are described. In addition, if one were to consider Linear Feedback Strategies for the finite horizon case i.e. the integrals in (2) were from 0 to a finite T, then the Linear Feedback Strategies –if they existed-they would have to be symmetric, as an inspection of the associated Ricatti –type equations would show).

$$u_i = L_1 x_i + L_2 z \tag{3}$$

where we set:

$$z = \frac{1}{M}(x_1 + x_2 + ... + x_M) \tag{4}$$

The $L_1(m \times n), L_2(m \times n)$ are constant matrices. This form is due to the symmetry of the state equation and the costs. The constants $L_1, L_2$ can be determined from the system of the $M$ coupled Riccati-type equations that characterize the Nash solution at hand (Refs 1,3,4). Another but equivalent way of determining them is the following: Let us consider the problem faced by player 1. He sees the state equations (1) for $i = 1, 2, ...M$, i.e.:

$$\frac{dx_1}{dt} = A_1 x_1 + \frac{A_2}{M}(x_1 + x_2 + ... + x_M) + B_1 u_1 + \frac{B_2}{M}(u_1 + u_2 + ... + u_M) \tag{5.1}$$

$$\frac{dx_2}{dt} = A_1 x_2 + \frac{A_2}{M}(x_1 + x_2 + ... + x_M) + B_1 u_2 + \frac{B_2}{M}(u_1 + u_2 + ... + u_M) \tag{5.2}$$

$$\frac{dx_3}{dt} = A_1 x_3 + \frac{A_2}{M}(x_1 + x_2 + ... + x_M) + B_1 u_3 + \frac{B_2}{M}(u_1 + u_2 + ... + u_M) \tag{5.3}$$

.    .
.    .
.    .

$$\frac{dx_M}{dt} = A_1 x_M + \frac{A_2}{M}(x_1 + x_2 + ... + x_M) + B_1 u_M + \frac{B_2}{M}(u_1 + u_2 + ... + u_M) \tag{5.M}$$

where he considers in (5.1)-(5.M):

$$u_i = L_1 x_i + L_2 z, \quad \text{for } i \neq 1, i = 2, 3, ...M \tag{6}$$

Adding up (5.1)-(5.M) yields:

$$\frac{d(x_1 + x_2 + ... + x_M)}{dt} = A_1(x_1 + x_2 + ... + x_M) +$$
$$+ A_2(x_1 + x_2 + ... + x_M) + B_1(u_1 + u_2 + ... + u_M) +$$
$$+ B_2(u_1 + u_2 + ... + u_M) =$$
$$= (A_1 + A_2)(x_1 + x_2 + ... + x_M) + (B_1 + B_2)(u_1 + u_2 + ... + u_M)$$

or

$$\frac{dz}{dt} = (A_1 + A_2)z + (B_1 + B_2)\frac{1}{M}(u_1 + u_2 + ... + u_M) \tag{7}$$

Adding up the equations (6) for $i = 2, 3, ... M$, yields:

$$u_2 + u_3 + ... + u_M = L_1(x_2 + x_3 + ... + x_M) + (M-1)L_2 z =$$
$$= L_1(Mz - x_1) + (M-1)L_2 z = \tag{8}$$
$$= -L_1 x_1 + [ML_1 + (M-1)L_2]z$$

Substituting $u_2 + u_3 + ... + u_M$ from (8) into (5.1) yields:

$$\frac{dx_1}{dt} = A_1 x_1 + \frac{A_2}{M}(x_1 + x_2 + ... + x_M) + B_1 u_1 + \frac{B_2}{M}(u_1 + u_2 + ... + u_M) =$$

$$= A_1 x_1 + A_2 z + (B_1 + \frac{B_2}{M})u_1 + \frac{B_2}{M}(u_2 + ... + u_M) =$$

$$= A_1 x_1 + A_2 z + (B_1 + \frac{B_2}{M})u_1 + \frac{B_2}{M}\{-L_1 x_1 + [ML_1 + (M-1)L_2]z\} =$$

$$= (A_1 - \frac{B_2}{M}L_1)x_1 + (A_2 + B_2 L_1 + B_2 \frac{M-1}{M}L_2)z + (B_1 + \frac{B_2}{M})u_1 =$$

Substituting $u_2 + u_3 + ... + u_M$ from (8) into (7) yields

$$\frac{dz}{dt} = (A_1 + A_2)z + (B_1 + B_2)\frac{1}{M}(u_1 + u_2 + ... + u_M) =$$

$$= (A_1 + A_2)z + (B_1 + B_2)\frac{1}{M}u_1 + (B_1 + B_2)\frac{1}{M}\{-L_1 x_1 + [ML_1 + (M-1)L_2]z\} =$$

$$= -(B_1 + B_2)\frac{1}{M}L_1 x_1 + [A_1 + A_2 + (B_1 + B_2)(L_1 + \frac{M-1}{M}L_2)]z + (B_1 + B_2)\frac{1}{M}u_1$$

or

$$\frac{d}{dt}\begin{bmatrix} x_1 \\ z \end{bmatrix} = A\begin{bmatrix} x_1 \\ z \end{bmatrix} + B u_1 \tag{9}$$

Where

$$A = \begin{bmatrix} A_1 - B_2 w L_1 & A_2 + B_2(L_1 + (1-w)L_2) \\ -(B_1 + B_2)w L_1 & A_1 + A_2 + (B_1 + B_2)(L_1 + (1-w)L_2) \end{bmatrix} =$$

$$= \begin{bmatrix} A_1 & A_2 \\ 0 & A_1 + A_2 \end{bmatrix} + \begin{bmatrix} 0 & B_2(L_1 + L_2) \\ 0 & (B_1 + B_2)(L_1 + L_2) \end{bmatrix} + w \begin{bmatrix} -B_2 L_1 & -B_2 L_2 \\ -(B_1 + B_2)L_1 & -(B_1 + B_2)L_2 \end{bmatrix} =$$

$$= \begin{bmatrix} A_1 & A_2 \\ 0 & A_1 + A_2 \end{bmatrix} + \begin{bmatrix} B_2 \\ B_1 + B_2 \end{bmatrix} \begin{bmatrix} 0 & L_1 + L_2 \end{bmatrix} - w \begin{bmatrix} B_2 \\ B_1 + B_2 \end{bmatrix} \begin{bmatrix} L_1 & L_2 \end{bmatrix} =$$

$$= \begin{bmatrix} A_1 & A_2 \\ 0 & A_1 + A_2 \end{bmatrix} + \begin{bmatrix} B_2 \\ B_1 + B_2 \end{bmatrix} \begin{bmatrix} L_1 & L_2 \end{bmatrix} \begin{bmatrix} 0 & I \\ 0 & I \end{bmatrix} - w \begin{bmatrix} B_2 \\ B_1 + B_2 \end{bmatrix} \begin{bmatrix} L_1 & L_2 \end{bmatrix} =$$

$$B = \begin{bmatrix} B_1 + B_2 w \\ (B_1 + B_2)w \end{bmatrix}$$

(10)

$$w = \frac{1}{M} \tag{11}$$

It is clear now that the problem of player 1 is to minimize his cost subject to the equations (9). The reason is that the state $x_1$ and the cost $J_1$ of the player 1 are influenced by $x_1, u_1$ and $z$ on which $z$ the total influence of $u_1$ is through (7). The solution is
given by the formula:

$$u_1 = -\begin{bmatrix} (B_1 + B_2 w)^T & (B_1 + B_2)^T w \end{bmatrix} K \begin{bmatrix} x_i \\ z \end{bmatrix} \tag{12}$$

Where

$$K = \begin{bmatrix} K_1 & K \\ K^T & K_2 \end{bmatrix} \tag{13}$$

is the solution of the matrix Ricatti equation:

$$0 = KA + A^T K + \tilde{Q} - K \begin{bmatrix} B_1 + B_2 w \\ (B_1 + B_2)w \end{bmatrix} \begin{bmatrix} (B_1 + B_2 w)^T & (B_1 + B_2)^T w \end{bmatrix} K, \tilde{Q} = \begin{bmatrix} Q & 0 \\ 0 & 0 \end{bmatrix} \tag{14}$$

The $L_1, L_2$ that are present in the Ricatti equation (14), through the matrix $A$ of (10) are to be identified with

$$\begin{bmatrix} L_1 & L_2 \end{bmatrix} = -\begin{bmatrix} (B_1 + B_2 w)^T & (B_1 + B_2)^T w \end{bmatrix} K \tag{15}$$

Substituting in (10) the $L_1, L_2$, by their equals using (15) we obtain for $A$ the equivalent form (16) that we denote by $A(K, w)$ in order to emphasize the dependence on $K$ and $w$:

$$A(K,w) = \begin{bmatrix} A_1 & A_2 \\ 0 & A_1 + A_2 \end{bmatrix} + \begin{bmatrix} B_2 \\ B_1 + B_2 \end{bmatrix} [L_1 \; L_2] \begin{bmatrix} 0 & I \\ 0 & I \end{bmatrix} - w \begin{bmatrix} B_2 \\ B_1 + B_2 \end{bmatrix} [L_1 \; L_2] =$$

$$= \begin{bmatrix} A_1 & A_2 \\ 0 & A_1 + A_2 \end{bmatrix} - \begin{bmatrix} B_2 \\ B_1 + B_2 \end{bmatrix} [(B_1 + B_2 w)^T \; (B_1 + B_2)^T w] K \begin{bmatrix} 0 & I \\ 0 & I \end{bmatrix} + w \begin{bmatrix} B_2 \\ B_1 + B_2 \end{bmatrix} [(B_1 + B_2 w)^T \; (B_1 + B_2)^T w] K$$

$$= \begin{bmatrix} A_1 & A_2 \\ 0 & A_1 + A_2 \end{bmatrix} - \begin{bmatrix} B_2 \\ B_1 + B_2 \end{bmatrix} [B_1^T \; 0] K \begin{bmatrix} 0 & I \\ 0 & I \end{bmatrix} - w \begin{bmatrix} B_2 \\ B_1 + B_2 \end{bmatrix} [B_2^T \; (B_1 + B_2)^T] K \begin{bmatrix} 0 & I \\ 0 & I \end{bmatrix} +$$

$$+ w \begin{bmatrix} B_2 \\ B_1 + B_2 \end{bmatrix} [B_1^T \; 0] K + w^2 \begin{bmatrix} B_2 \\ B_1 + B_2 \end{bmatrix} [B_2^T \; (B_1 + B_2)^T] K$$

(16)

It is this $A(K,w)$ of (16) that is used in (14). Notice that (14) has several quadratic terms in $K$ besides the last one appearing in (14), since $A(K,w)$ itself is a linear function of $K$. The solution of (14) if it exists, is a function $K(w)$ which has the value $K(0)$ for $w = 0$. Since we are going to let $M \to \infty$, we will study the behavior of the Ricatti equation (14) by allowing $w$ to be a continuous variable close to 0.

Let us set:

$$R(K,w) = KA(K,w) + A^T(K,w)K + \tilde{Q} - K \begin{bmatrix} B_1 + B_2 w \\ (B_1 + B_2)w \end{bmatrix} [(B_1 + B_2 w)^T \; (B_1 + B_2)^T w] K \qquad (17)$$

The function $R(K,w)$ is analytic in its arguments. Let us set

$$K(w) = K_0 + \overline{K}_1 w + \overline{K}_2 w^2 + \ldots \qquad (18)$$

We plug $K(w) = K_0 + \overline{K}_1 w + \overline{K}_2 w^2 + \ldots$ in (17) and group together the terms corresponding to the same powers of $w$ to obtain

$$R(K(w),w) = K(w)A(K(w),w) + A^T(K(w),w)K(w) + \tilde{Q} - K(w) \begin{bmatrix} B_1 + B_2 w \\ (B_1 + B_2)w \end{bmatrix} [(B_1 + B_2 w)^T \; (B_1 + B_2)^T w] K(w) =$$

$$= R_0(K_0) + wR_1(K_0, \overline{K}_1) + w^2 R_2(K_0, \overline{K}_1, \overline{K}_2) + \ldots + w^n R_n(K_0, \overline{K}_1, \ldots, \overline{K}_n) + \ldots \qquad (19)$$

Where

$$R_0(K_0) = K_0 \{ \begin{bmatrix} A_1 & A_2 \\ 0 & A_1 + A_2 \end{bmatrix} - \begin{bmatrix} B_2 B_1^T & 0 \\ (B_1 + B_2)B_1^T & 0 \end{bmatrix} K_0 \begin{bmatrix} 0 & I \\ 0 & I \end{bmatrix} \} + \{\ldots\}^T K_0 + \tilde{Q} - K_0 \begin{bmatrix} B_1 B_1^T & 0 \\ 0 & 0 \end{bmatrix} K_0 \qquad (20)$$

$$R_1(K_0, \bar{K}_1) = \bar{K}_1\{\begin{bmatrix} A_1 & A_2 \\ 0 & A_1 + A_2 \end{bmatrix} - \begin{bmatrix} B_2 B_1^T & 0 \\ (B_1 + B_2) B_1^T & 0 \end{bmatrix} K_0 \begin{bmatrix} 0 & I \\ 0 & I \end{bmatrix}\} + \{...\}^T \bar{K}_1 +$$

$$-K_0\{\begin{bmatrix} B_2 \\ B_1 + B_2 \end{bmatrix} \begin{bmatrix} B_1^T & 0 \end{bmatrix} \bar{K}_1 \begin{bmatrix} 0 & I \\ 0 & I \end{bmatrix}\} - \{...\}^T K_0 \qquad (21)$$

$$-K_0 \begin{bmatrix} B_1 B_1^T & 0 \\ 0 & 0 \end{bmatrix} \bar{K}_1 - \bar{K}_1 \begin{bmatrix} B_1 B_1^T & 0 \\ 0 & 0 \end{bmatrix} K_0 + \text{Quadratic terms of}(K_0)$$

Similarly we can derive the formulae for the other $R_n(K_0, \bar{K}_1, ..., \bar{K}_n)$'s.

Let us consider the matrix:

$$A_c(0) = \begin{bmatrix} A_1 & A_2 \\ 0 & A_1 + A_2 \end{bmatrix} - \begin{bmatrix} B_2 B_1^T & 0 \\ (B_1 + B_2) B_1^T & 0 \end{bmatrix} K_0 \begin{bmatrix} 0 & I \\ 0 & I \end{bmatrix} - \begin{bmatrix} B_1 B_1^T & 0 \\ 0 & 0 \end{bmatrix} K_0 =$$

$$\begin{bmatrix} A_1 & A_2 \\ 0 & A_1 + A_2 \end{bmatrix} - \begin{bmatrix} B_2 B_1^T & 0 \\ (B_1 + B_2) B_1^T & 0 \end{bmatrix} \begin{bmatrix} K_1 & K \\ K^T & K_2 \end{bmatrix} \begin{bmatrix} 0 & I \\ 0 & I \end{bmatrix} - \begin{bmatrix} B_1 B_1^T & 0 \\ 0 & 0 \end{bmatrix} \begin{bmatrix} K_1 & K \\ K^T & K_2 \end{bmatrix} \qquad (22)$$

$$= \begin{bmatrix} A_1 - B_1 B_1^T K_1 & A_2 - B_2 B_1^T (K_1 + K) - B_1 B_1^T K \\ 0 & A_1 + A_2 - (B_1 + B_2) B_1^T (K_1 + K) \end{bmatrix} = \begin{bmatrix} A_{c1} & A_{c0} \\ 0 & A_{c2} \end{bmatrix}$$

It holds:

$$\begin{aligned} A_{c1} &= A_1 - B_1 B_1^T K_1 \\ A_{c2} &= A_1 + A_2 - (B_1 + B_2) B_1^T (K_1 + K) \\ A_{c0} &= A_2 - B_2 B_1^T (K_1 + K) - B_1 B_1^T K \\ A_{c0} &= A_{c2} - A_{c1} \end{aligned} \qquad (23)$$

Then

$$R_0(K_0) = K_0 \begin{bmatrix} A_1 - B_1 B_1^T K_1 & A_2 - B_2 B_1^T (K_1 + K) - B_1 B_1^T K \\ 0 & A_1 + A_2 - (B_1 + B_2) B_1^T (K_1 + K) \end{bmatrix} + [..]^T K_0 + \tilde{Q} + K_0 \begin{bmatrix} B_1 B_1^T & 0 \\ 0 & 0 \end{bmatrix} K_0 =$$

$$= \begin{bmatrix} K_1 & K \\ K^T & K_2 \end{bmatrix} \begin{bmatrix} A_1 - B_1 B_1^T K_1 & A_2 - B_2 B_1^T (K_1 + K) - B_1 B_1^T K \\ 0 & A_1 + A_2 - (B_1 + B_2) B_1^T (K_1 + K) \end{bmatrix} + [..]^T K_0 + \tilde{Q} + K_0 \begin{bmatrix} B_1 B_1^T & 0 \\ 0 & 0 \end{bmatrix} K_0 =$$

$$= \begin{bmatrix} K_1(A_1 - B_1 B_1^T K_1) & K_1(A_2 - B_2 B_1^T (K_1 + K) - B_1 B_1^T K) + K(A_1 + A_2 - (B_1 + B_2) B_1^T (K_1 + K)) \\ K^T(A_1 - B_1 B_1^T K_1) & K^T(A_2 - B_2 B_1^T (K_1 + K) - B_1 B_1^T K) + K_2(A_1 + A_2 - (B_1 + B_2) B_1^T (K_1 + K)) \end{bmatrix} + [...]^T +$$

$$+ \begin{bmatrix} Q + K_1 B_1 B_1^T K_1 & K_1 B_1 B_1^T K \\ K^T B_1 B_1^T K_1 & K^T B_1 B_1^T K \end{bmatrix}$$

$$(24)$$

Let us introduce the operator $L(K_0, X)$ which is actually the derivative of $R(K, w)$ with respect to $K$ calculated at $w = 0$. For

$$X = \begin{bmatrix} \chi_1 & \chi \\ \chi^T & \chi_2 \end{bmatrix} \quad (25)$$

we define:

$$L(K_0, X) = X\left\{ \begin{bmatrix} A_1 & A_2 \\ 0 & A_1 + A_2 \end{bmatrix} - \begin{bmatrix} B_2 B_1^T & 0 \\ (B_1 + B_2)B_1^T & 0 \end{bmatrix} K_0 \begin{bmatrix} 0 & I \\ 0 & I \end{bmatrix} - \begin{bmatrix} B_1 B_1^T & 0 \\ 0 & 0 \end{bmatrix} K_0 \right\} + \{...\}^T X +$$

$$- K_0 \begin{bmatrix} B_2 \\ B_1 + B_2 \end{bmatrix} \begin{bmatrix} B_1^T & 0 \end{bmatrix} X \begin{bmatrix} 0 & I \\ 0 & I \end{bmatrix} - \left( K_0 \begin{bmatrix} B_2 \\ B_1 + B_2 \end{bmatrix} \begin{bmatrix} B_1^T & 0 \end{bmatrix} X \begin{bmatrix} 0 & I \\ 0 & I \end{bmatrix} \right)^T =$$

$$= X A_c(0) + A_0^T(0) X +$$

$$- \begin{bmatrix} 0 & (K_1 B_2 B_1^T + K(B_1 + B_2)B_1^T)(\chi_1 + \chi) \\ 0 & (K^T B_2 B_1^T + K_2 (B_1 + B_2)B_1^T)(\chi_1 + \chi) \end{bmatrix} - \begin{bmatrix} 0 & (K_1 B_2 B_1^T + K(B_1 + B_2)B_1^T)(\chi_1 + \chi) \\ 0 & (K^T B_2 B_1^T + K_2 (B_1 + B_2)B_1^T)(\chi_1 + \chi) \end{bmatrix}^T =$$

$$= \begin{bmatrix} \chi_1 & \chi \\ \chi^T & \chi_2 \end{bmatrix} \begin{bmatrix} A_{c1} & A_{c0} \\ 0 & A_{c2} \end{bmatrix} + A_0^T(0) X - \begin{bmatrix} 0 & (K_1 B_2 B_1^T + K(B_1 + B_2)B_1^T)(\chi_1 + \chi) \\ 0 & (K^T B_2 B_1^T + K_2 (B_1 + B_2)B_1^T)(\chi_1 + \chi) \end{bmatrix} - \begin{bmatrix} 0 & (K_1 B_2 B_1^T + K(B_1 + B_2)B_1^T)(x_1 + x) \\ 0 & (K^T B_2 B_1^T + K_2 (B_1 + B_2)B_1^T)(x_1 + x) \end{bmatrix} =$$

$$= \begin{bmatrix} \chi_1 A_{c1} & \chi_1 A_{c0} + \chi A_{c2} - (K_1 B_2 B_1^T + K(B_1 + B_2)B_1^T)(\chi_1 + \chi) \\ \chi^T A_{c1} & \chi^T A_{c0} + \chi_2 A_{c2} - (K^T B_2 B_1^T + K_2 (B_1 + B_2)B_1^T)(\chi_1 + \chi) \end{bmatrix} + [...]^T = \begin{bmatrix} L_{11}(K_0, X) & L_{12}(K_0, X) \\ L_{21}(K_0, X) & L_{22}(K_0, X) \end{bmatrix}$$

(26)

Equivalently, the operator $L(K_0, X)$ can be written as a "matrix operator" multiplying a "vector":

$$\begin{bmatrix} L_{11}(K_0, X) \\ L_{12}(K_0, X) \\ L_{22}(K_0, X) \end{bmatrix} = \begin{bmatrix} \chi_1 A_{c1} + A^T{}_{c1} \chi_1 & 0 & 0 \\ L_{21}(\chi_1) & \chi A_{c2} + (A_1^T - Y(B_1 + B_2)B_1^T)\chi & 0 \\ L_{31}(\chi_1) & L_{32}(\chi) & \chi_2 A_{c2} + A^T{}_{c2} \chi_2 \end{bmatrix} \quad (27)$$

which makes the invertibility study of $L(K_0, X)$ quite transparent.
We can now state the following proposition:

**Proposition 1**
If at $w = 0$ the equation $R(K, 0) = 0$ has a solution $K_0$, i.e. $R(K_0, 0) = 0$ and if the operator $L(K_0, X)$ as a linear operator on $X$ is invertible, then in some neighborhood of $w = 0$ the equation $R(K, w) = 0$ has a unique solution $K(w)$ which is an analytic function of $w$ and has an expansion (18), where the $K_0, \overline{K}_1, \overline{K}_2,...$ are the unique solutions of $R_0(K_0) = 0, R_1(K_0, \overline{K}_1) = 0, R_2(K_0, \overline{K}_1, \overline{K}_2) = 0,...$

**Proof**
The proof of this proposition is a straightforward application of the implicit function theorem for analytic functions (see Theorem 8.6 in Ref.20) and uses the formulae already developed in (17)-(27).
●

It is important to notice that if we want to find the $\overline{K}_n$'s for $n \geq 1$ we have to find the coefficient of the power $w^n$ in the equation $R(K(w), w) = 0$ and set it equal to 0, which yields an equation linear in $\overline{K}_n$ of the form:

$L(K_0, \bar{K}_n) = F(K_0, \bar{K}_1, ..., \bar{K}_{n-1})$, where the linear operator $L(K_0, \bar{K}_n)$ is the one defined in (26) and $F(K_0, \bar{K}_1, ..., \bar{K}_{n-1})$ is a nonlinear function containing multiplicative terms of its arguments.

Let us now introduce the closed loop matrix $A_c(w)$ by the formula:

$$A_c(w) = A(K, w) = \begin{bmatrix} A_1 & A_2 \\ 0 & A_1 + A_2 \end{bmatrix} - \begin{bmatrix} B_2 \\ B_1 + B_2 \end{bmatrix} \begin{bmatrix} (B_1 + B_2 w)^T & (B_1 + B_2)^T w \end{bmatrix} K(w) \begin{bmatrix} -w & 1 \\ w^2 & 1 - 2w \end{bmatrix} +$$
$$- \begin{bmatrix} B_1 + B_2 w \\ (B_1 + B_2)w \end{bmatrix} \begin{bmatrix} (B_1 + B_2 w)^T & (B_1 + B_2)^T w \end{bmatrix} K(w) \tag{28}$$

This is the closed loop matrix of (9) that results when all the players use their optimal strategies and we will have:

$$\frac{d}{dt} \begin{bmatrix} x_i \\ z \end{bmatrix} = A_c(w) \begin{bmatrix} x_i \\ z \end{bmatrix} \tag{29}$$

We can now state the second proposition that pertains to the existence of a Nash equilibrium.

**Proposition 2**
**2.1** If the Ricatti equation $R(K, w) = 0$ has a solution $K(w) \geq 0$ which yields an asymptotically stable closed loop matrix $A_c(w)$, then the resulting $u_1^*$ of (15) and the similar $u_2^*, ..., u_M^*$ are in Nash equilibrium.
**2.2** If the Ricatti equation $R_0(K_0) = 0$, where $R_0(K_0)$ is given in (20) has a solution $K_0 \geq 0$ and the matrix $A_c(0)$ of (22) (or from (28) with $w = 0$) is asymptotically stable, and the operator $L(K_0, X)$ is invertible, then for $M$ sufficiently large, all the games considered have a Nash equilibrium with asymptotically stable closed loop matrices $A_c(w)$ (which are as in(28)) and tend to $A_c(0)$ as $M \to \infty$. The corresponding $K(w)$ which determines the $u_i^*$ can be approximated up to order one in $w$ by $K_0 + K_1 w$, where $K_0, K_1$ solve the two equations $R_0(K_0) = 0, R_1(K_0, K_1) = 0$.
**2.3** The $L_1, L_2$ that are calculated by using (15) with $K_0$ and $w = 0$ constitute an $\varepsilon(M, x(0))$-Nash equilibrium.

**Proof**
The proof of this Proposition is an immediate consequence of the previously presented analysis. ●

**Comment 1**
The equation $R_0(K_0) = 0$ results in the system:
$$0 = K_1 A_1 + A_1^T K_1 + Q - K_1 B_1 B_1^T K_1$$
$$0 = K_1(A_2 - B_2 B_1^T(K_1 + K) - B_1 B_1^T K) + K(A_1 + A_2 - (B_1 + B_2)B_1^T(K_1 + K)) + (A_1 - B_1 B_1^T K_1)^T K + K_1 B_1 B_1^T K$$
$$0 = K^T(A_2 - B_2 B_1^T(K_1 + K) - B_1 B_1^T K) + (A_2 - B_2 B_1^T(K_1 + K) - B_1 B_1^T K)^T K$$
$$+ K_2(A_1 + A_2 - (B_1 + B_2)B_1^T(K_1 + K)) + (A_1 + A_2 - (B_1 + B_2)B_1^T(K_1 + K))^T K_2$$

Substituting the second one with the sum of the first two we have the following equivalent system that $K_1, K_1 + K, K_2$ have to satisfy:

$0 = K_1 A_1 + A_1^T K_1 + Q - K_1 B_1 B_1^T K_1$

$0 = (K_1 + K)(A_2 + A_1) + A_1^T(K + K_1) - (K + K_1)(B_1 + B_2) B_1^T (K + K_1) + Q$

$0 = K^T (A_2 - B_2 B_1^T (K_1 + K) - B_1 B_1^T K) + (A_2 - B_2 B_1^T (K_1 + K) - B_1 B_1^T K)^T K$

$+ K_2 (A_1 + A_2 - (B_1 + B_2) B_1^T (K_1 + K)) + (A_1 + A_2 - (B_1 + B_2) B_1^T (K_1 + K))^T K_2$

or

$$0 = K_1 A_1 + A_1^T K_1 + Q - K_1 B_1 B_1^T K_1 \qquad (30)$$

$$0 = Y(A_2 + A_1) + A_1^T Y - Y(B_1 + B_2) B_1^T Y + Q \qquad (31)$$

$$K = Y - K_1$$

$$0 = K_2 A_{c2} + A_{c2}^T K_2 + K^T A_{c0} + A_{c0}^T K \qquad (32)$$

Where

$$A_{c1} = A_1 - B_1 B_1^T K_1$$

$$A_{c2} = A_1 + A_2 - (B_1 + B_2) B_1^T Y \qquad (33)$$

$$A_{c0} = A_{c2} - A_{c1}$$

It is clear that the gain $K_1$ that multiplies the state $x_i$ of player i depends only on his part of the system and cost. The Ricatti equation (30) for $K_1$ is the classical one and the $K_1$ exists and is positive definite under the usual assumptions. The gain $K$ that couples the controller of the player 1 with the states of the others may fail to exist since its existence depends on the generalized Ricatti (31). Notice that the study of (31) is reduced equivalently to the study of the "Perturbed Hamiltonian"

$$H = \begin{bmatrix} A_1 + A_2 & -(B_1 + B_2) B_1^T \\ -Q & -A_1^T \end{bmatrix} = \begin{bmatrix} A_1 & -B_1 B_1^T \\ -Q & -A_1^T \end{bmatrix} + \begin{bmatrix} A_2 & -B_2 B_1^T \\ 0 & 0 \end{bmatrix} \qquad (34)$$

If $Y$ satisfies (31), then:

$$\begin{bmatrix} I & 0 \\ -Y & I \end{bmatrix} \begin{bmatrix} A_1 + A_2 & -(B_1 + B_2) B_1^T \\ -Q & -A_1^T \end{bmatrix} \begin{bmatrix} I & 0 \\ Y & I \end{bmatrix} = \begin{bmatrix} A_1 + A_2 - (B_1 + B_2) B_1^T Y & -(B_1 + B_2) B_1^T \\ -Y(A_1 + A_2) - A_1^T Y + Y(B_1 + B_2) B_1^T Y - Q & -A_1^T + Y(B_1 + B_2) B_1^T \end{bmatrix} =$$

$$= \begin{bmatrix} A_1 + A_2 - (B_1 + B_2) B_1^T Y & -(B_1 + B_2) B_1^T \\ 0 & -A_1^T + Y(B_1 + B_2) B_1^T \end{bmatrix}$$

(35)

and thus the matrices $A_c = A_1 + A_2 - (B_1 + B_2)^T BY$ , $-(A_1^T - Y(B_1 + B_2) B_1^T))$ have the eigenvalues of $H$ and can be calculated by using the corresponding eigenvectors of $H$, see Ref.19.

Notice that in order for the Nash game to have a solution we need both $A_{c1}, A_{c2}$ to be asymptotically stable, a fact that will hold for $A_{c1}$ if $A_1, B_1$ is a controllable pair. The conditions for the asymptotic stability of $A_{c2}$ are less obvious and depend on the eigenstructure of $H$. Actually there are cases where there is no solution resulting in $A_{c2}$ asymptotically stable, or there can be more than one solution that yield $A_{c2}$ asymptotically stable which would mean correspondingly that we have no or many Nash equilibria. Finally notice that solving for $K_2$ is a linear problem that always has a solution if $A_{c2}$ is asymptotically stable. Thus the existence of a Nash solution amounts to studying (30) and (31) and demanding that both $A_{c1}, A_{c2}$ are asymptotically stable.

**Comment 2**
The invertibility of $L(K_0, X)$ is equivalent to having invertibility of the operators:

$$\bar{L}_{11}(\chi_1) = \chi_1 A_{c1} + A^T_{c1} \chi_1$$
$$\bar{L}_{22}(\chi) = \chi A_{c2} + (A_{c1} - B_1 B_2^T K_1 - B_1(B_1 + B_2)^T K^T)^T \chi \qquad (36)$$
$$\bar{L}_{33}(\chi_2) = \chi_2 A_{c2} + A^T_{c2} \chi_2$$

The first and third one are invertible if $A_{c1}, A_{c2}$ are asymptotically stable. It is the invertibility of the second one that is the more interesting. Let us look at it more carefully:

$$\chi A_{c2} + (A_1 - B_1(B_1 + B_2)^T (K_1 + K^T))^T \chi =$$
$$= \chi(A_1 + A_2 - (B_1 + B_2) B_1^T (K_1 + K)) + (A_1^T - (K_1 + K)(B_1 + B_2) B_1^T) \chi \qquad (37)$$

This equation involves the matrices $A_{c2} = A_1 + A_2 - (B_1 + B_2)^T BY$ and $-(A_1^T - Y(B_1 + B_2) B_1^T)$ that appear in (31) (or (35)), and thus we have invertibility of $L(K_0, X)$ if and only if:

$$eigenvalue A_{c2} + eigenvalue(A_1^T - (K_1 + K)(B_1 + B_2) B_1^T)) \neq 0 \qquad (38)$$

Notice also that equation (37) is actually the perturbation of the equation that (31) that gives $Y = K_1 + K$. To see that, let us perturb the solution $Y$ of (31) to $Y + \Delta$ to get:

$$(Y + \Delta)(A_2 + A_1) + A_1^T(Y + \Delta) - (Y + \Delta)(B_1 + B_2) B_1^T (Y + \Delta) + Q =$$
$$= Y(A_2 + A_1) + A_1^T Y - Y(B_1 + B_2) B_1^T Y + Q + \Delta(A_1 + A_2 - (B_1 + B_2) B_1^T (K_1 + K)) + (A_1^T - (K_1 + K)(B_1 + B_2) B_1^T)\Delta +$$
$$-\Delta(B_1 + B_2) B_1^T \Delta =$$
$$= \Delta(A_1 + A_2 - (B_1 + B_2) B_1^T (K_1 + K)) + (A_1^T - (K_1 + K)(B_1 + B_2) B_1^T)\Delta - \Delta(B_1 + B_2) B_1^T \Delta$$
$$(39)$$

Therefore the condition for invertibility of $L(K_0, X)$ is equivalent to asking that the perturbation of the equation that determines $Y = K_1 + K$ has no zero eigenvalue. If we demand more than that, namely that all the perturbation eigenvalues of the perturbed equation for $Y = K_1 + K$ are negative, i.e.:
$$eigenvalue(A_1 + A_2 - (B_1 + B_2) B_1^T (K_1 + K)) + eigenvalue(A_1^T - (K_1 + K)(B_1 + B_2) B_1^T)) < 0 \qquad (40)$$
or

$$eigenvalue A_{c2} + eigenvalue(A_1^T - (K_1 + K)(B_1 + B_2)B_1^T)) < 0$$

this would imply a kind of stability of the solutions of (31) that is beneficial for any algorithm that solves (31). Actually this is equivalent to something more interesting, namely it is equivalent to asking that the infinite time solutions, i.e. the $L_i^*$ gains, can be approached as stable limits of the gains of the finite time horizon Nash game. Such a Nash equilibrium we will call "Stable Nash equilibrium". (We remind the reader that a solution of the infinite horizon problem is not necessarily a limit of the finite horizon solution. In relation to that and the interplay between finite time horizon and Infinite Time horizon solutions, seeRefs.17,18, 3, 4). The justification of this claim follows. Consider the cost:

$$J_i = \frac{1}{2} x_i^T(t_f) Q_f x_i(t_f) + \frac{1}{2} \int_0^{t_f} (x_i^T(t) Q x_i(t) + u_i^T(t) u_i(t)) dt \tag{41}$$

for some $Q_f \geq 0$.

The state equations remain the same and all the transformations in (1)-(12) hold.
The finite time Ricatti equation yields a time varying solution which satisfies:

$$-\frac{dK(w,t)}{dt} = R(K(w,t), w), \quad K(w, t_f) = \begin{bmatrix} Q_f & 0 \\ 0 & 0 \end{bmatrix}$$

Let

$$K(w,t) = \begin{bmatrix} K_1(w,t) & K(w,t) \\ K(w,t) & K_2(w,t) \end{bmatrix}$$

Assuming $w = 0$ we see that the following differential equations are obtained:

$$-\frac{dK_1(t)}{dt} = K_1(t) A_1 + A_1^T K_1(t) + Q - K_1(t) B_1 B_1^T K_1(t), K_1(t_f) = Q_f \tag{42a}$$

$$-\frac{dY(t)}{dt} = Y(t)(A_2 + A_1) + A_1^T Y(t) - Y(t)(B_1 + B_2) B_1^T Y(t) + Q_1, Y(t_f) = 0 \tag{42b}$$

$$K(t) = Y(t) - K_1(t)$$

$$-\frac{dK_2(t)}{dt} = K_2(t) A_{c2} + A_{c2}^T K_2(t) + K^T(t) A_{c0} + A_{c0}^T K(t) + K^T(t) B_1 B_1^T K(t), K(t_f) = 0 \tag{42c}$$

$$A_{c2}(t) = A_1 + A_2 - (B_1 + B_2) B_1^T Y(t))$$

The solution of (42a) goes to the positive definite stabilizing solution of (30) under the usual controllability assumptions on $A_1, B_1$. The differential equation (42b) has as equilibrium point the solution $Y$ of (31). Linearizing (42b) around this $Y$ we get the linearized equation (39), and thus we conclude that the $Y$ is an asymptotically stable equilibrium of (42b) if and only if: $eigenvalue A_{c2} + eigenvalue(A_1^T - (K_1 + K)(B_1 + B_2)B_1^T)) < 0$. Therefore, if

$eigenvalue A_{c1} < 0$ and $eigenvalue A_{c2} < 0$ and $eigenvalue A_{c2} + eigenvalue(A_1^T - (K_1 + K)(B_1 + B_2)B_1^T)) < 0$

we will have a Nash equilibrium of the infinite horizon LQ game that can be also considered "Stable" in the sense that is the limit of a finite time horizon equilibrium.
Let us now summarize the material of Comments 2 and 3 in the form of a Proposition.

**Proposition 3**

**3.1** The Ricatti equation $R_0(K_0) = 0$ or equivalently the system (30)-(32) has a solution that results in asymptotically stable $A_{c1}, A_{c2}$ if the pair $A_1, B_1$ is completely controllable and if the equation (31) has a solution $Y$ that makes the $A_{c2}$ asymptotically stable. This last requirement is equivalent to asking that the $H$ matrix has n negative eigenvalues $\lambda_1, \lambda_1, ..., \lambda_n$ whose eigenvectors have their upper n-dimensional parts linearly independent, and the other n eigenvalues $\lambda_{n+1}, \lambda_{n+2}, ..., \lambda_{2n}$ satisfy: $\lambda_i - \lambda_j \neq 0, i = 1, 2, ..., n, j = n+1, n+2, ..., 2n$. The eigenvectors corresponding to $\lambda_1, \lambda_1, ..., \lambda_n$ are used to construct $Y$ (see Ref.19).

**3.2** If the assumption of 3.1 hold and in addition the eigenvalues satisfy: $\lambda_i < \lambda_j, i = 1, 2, ..., n, j = n+1, n+2, ..., 2n$, then the Nash Equilibrium is a "Stable" one in the sense delineated above.

**Proof**
The proof is actually given in Comments 2 and 3. We use the classical construction of Ref.19. As evidenced from (37), $H$ has the eigenvalues of $A_{c2}$ and the negative eigenvalues of $A_1^T - (K_1 + K)(B_1 + B_2)B_1^T = A_1^T - Y(B_1 + B_2)B_1^T$.

●

**Comment 3**
The study of $H$ is quite central to the existence and character of the Nash solution and as such it merits independent investigation. First of all it is clear from (34) that $H$ is equal to the classical Hamiltonian which corresponds to the classical Ricatti (30), and since the eigenvalues are continuous functions of the matrix entries, if we think of $A_2, B_2$ as perturbations, then for sufficiently small values of them the assumptions and thus the conclusions of Proposition 3 hold. Thus it is easy to produce sufficiency nonempty conditions for Proposition 3 to hold. Notice also that $H$ is not Hamiltonian in the sense encountered in the Linear Quadratic Control Theory and the study of the classical Ricatti equation, see Refs.3, 4, and 19 and it can be any arbitrary $2n \times 2n$ matrix as the choices of $A_1, A_2, B_1, B_2$ can produce any value for the terms they determine in $H$. The $Q$ term of $H$ seems to introduce a restriction since it is symmetric and positive semidefinite. Actually any quadratic matrix equation of the form $H_{11}Y + YH_{12} + YH_{22}Y + H_{21} = 0$ can be transformed to having $H_{21}$ symmetric and positive semidefinite by using $H_{21} = UDV^T$ the singular value decomposition of $H_{12}$, with $D$ diagonal and positive semidefinite. Pre- and post-multiplying the equation with $U^T$ and $V$ respectively and considering as new unknown the $\bar{Y} = U^T Y V$, we have a new quadratic matrix equation with symmetric positive semidefinite constant term. Thus the $Q$ term does not really provide any structure and thus the $H$ matrix we deal with here does not have any particular structure.

**Comment 4**
The underlying theme of this work is the presence of very many symmetric players, and therefore the limiting behavior as the number of players $M$ grows to infinity is considered. It should be noticed that nowhere did we refer to a game with an infinite number of players, or to $K_0$ as defining the Nash equilibrium strategy of a player in the presence of an infinite number of players. Nonetheless if one were to make sense of such a limit per se-as if the players are infinite in multitude, the equations (9) are of importance. Notice by the way, that we derived (9) where the variable $z$ concatenates the influence of all the other players on player one, and derived then the Nash equilibrium, although we could have derived the Nash equilibrium working directly on equations (5.1)-(5.$M$). Let us look more carefully at equations (9), restated below for convenience.
If we consider that the $L_1, L_2$ have values that converge to definite limits as $M \to \infty$, a fact that holds under the assumptions of Proposition 2, then the system (9) in the limit $M \to \infty$ behaves like:

$$\frac{dx_1}{dt} = A_1 x_1 + (A_2 + L_1 + L_2)z + u_1, u_1 = -B_1 B_1^T K_1 x_1 \tag{43}$$

$$\frac{dz}{dt} = (A_1 + A_2 - (B_1 + B_2)B_1^T(K_1 + K))z \tag{44}$$

Equivalently we can say that player one faces a control problem of minimizing $J_1$ (as in (2)), where his control is $u_1$, and his state $x_1$ is influenced by a state variable $z$ which $z$ is available to him but is not at all influenced by him: The state $z$ obeys the evolution equation (44) which is not at all influenced by $x_1$ or $u_1$, as should be expected, since one player on his own should not have any influence on the collective behavior of a infinite number of fellow players. The question is whether we can think of the equation (44) as resulting from a control problem with state equation:

$$\frac{dz_e}{dt} = (A_1 + A_2)z_e + (B_1 + B_2)u_e \tag{45}$$

and cost to be minimized

$$\min J_e = \frac{1}{2}(\int_0^\infty (z_e(t)^T Q_e z_e(t) + 2u_e(t)^T S_e z_e(t) + u_e(t)^T u_e(t))dt \tag{46}$$

which for appropriate choices of $Q_e$ $S_e$, has as solution:

$$u_e = -B_1^T(K_1 + K))z_e \tag{47}$$

(Notice that is a particular case of an "inverse" problem" and such problems have been studied in the past in more generality, see for example:Refs 21,22 ). The state $z_e$ and the control $u_e$, can be thought of as the collective state and control of the other infinite in multitude players that determine the behavior of the "market" faced by player one.

Of course it is assumed that:

$$\begin{bmatrix} Q_e & S_e^T \\ S_e & I \end{bmatrix} \geq 0 \Leftrightarrow Q_e - S_e^T S_e \geq 0$$

The solution of this problem is:

$$u_e = -[(B_1 + B_2)^T P + S_e]z_e$$

where P solves the Ricatti equation:

$$0 = P(A_1 + A_2 - (B_1 + B_2)^T S_e) + (A_1 + A_2 - (B_1 + B_2)^T S_e)^T P + Q_e - S_e^T S_e - P(B_1 + B_2)(B_1 + B_2)^T P$$

$Q_e, S_e$ have to be chosen so that the optimal control solution (47) has the value $u_e = -B_1^T(K_1 + K))z_e$, i.e.

$$(B_1 + B_2)^T P + S_e = B_1^T(K_1 + K)$$

or

$$S_e = B_1^T Y - (B_1 + B_2)^T P$$

Substituting $S_e$ with it's equal from above we get:

$$0 = P(A_1 + A_2) + (A_1 + A_2)^T P + Q_e - (B_1^T Y)^T B_1^T Y$$

Therefore in order that the problem (45)-(46) has the solution (47), we should choose:

$$Q_e = Q_e^T \geq 0, S_e, Q_e - S_e^T S_e \geq 0, P = P^T \geq 0 \tag{48}$$

so that the following hold:

$$S_e = B_1^T Y - (B_1 + B_2)^T P \tag{49}$$

$$Q_e = -P(A_1 + A_2) - (A_1 + A_2)^T P + (B_1^T Y)^T B_1^T Y \geq 0 \qquad (50)$$

where $Y$ solves

$$0 = Y(A_2 + A_1) + A_1^T Y - Y(B_1 + B_2)B_1^T Y + Q \qquad (51)$$

These conditions give $Q_e, S_e, P$ as functions of the parameters $A_1, A_2, B_1, B_2, Q$ and they can always be satisfied as the following choice shows:

$$P = 0, S_e = B_1^T Y, \quad Q_e = (B_1^T Y)^T B_1^T Y \geq 0$$

With this choice the cost (46) is:

$$\min J_e = \frac{1}{2} \int_0^\infty \begin{bmatrix} z_e(t) \\ u_e(t) \end{bmatrix}^T \begin{bmatrix} Y^T B_1 B_1^T Y_1 & Y^T B_1 \\ B_1^T Y & I \end{bmatrix} \begin{bmatrix} z_e(t) \\ u_e(t) \end{bmatrix} dt$$

which results in optimal value 0 since $u_e = -B_1^T(K_1 + K))z_e = -S_e z_e$

A special possible choice is to have $S_e = 0$ and then finding $P, Q_e$ may be or not be possible. For example, let $B_2 = 0$ so that we have coupling of the state equations of the players only through the $A_2$ term. We can take $S_e = 0, Y = P$, if $Y = Y^T$, which yields:

$$Q_e = -Y(A_1 + A_2) - (A_1 + A_2)^T Y + Y^T B_1 B_1^T Y \geq 0$$

$$Q = -Y(A_2 + A_1) - A_1^T Y + YB_1 B_1^T Y$$

or

$$Q_e = Q + A_2 Y$$

The case $Y = Y^T$ can occur for example if $A_2^T Y = YA_2^T$ (a fact that cannot be guaranteed a priori) since then the equation for $Y$ can be written as

$$0 = Y(\frac{1}{2}A_2 + A_1) + (\frac{1}{2}A_2 + A_1)^T Y + Q - YB_1 B_1^T Y$$

and obviously has a symmetric solution. For example if the coupling matrix $A_2 = a_2 I$ with $a_2$ a real scalar, this is the case and $Q_e = Q + A_2 Y = Q + a_2 Y$ is acceptable if it is positive semidefinite.

**Comment 5**

It would be important to consider whether $K_1 \geq 0, K_1 \leq 0,$ or $K_1$ indefinite. For example, if it is positive definite, this means that as the number of players goes to infinity the cost of each player at the Nash equilibrium is decreasing. Let us consider the equation:

$$R_1(K_0, \bar{K}_1) = 0$$

which can be written as:

$$\begin{bmatrix} L_{11}(K_0, \bar{K}_1) \\ L_{12}(K_0, \bar{K}_1) \\ L_{22}(K_0, \bar{K}_1) \end{bmatrix} = \begin{bmatrix} \chi_1 A_{c1} + A_{c1}^T \chi_1 & 0 & 0 \\ L_{21}(\chi_1) & \chi A_{c2} + (A_1^T - YB_1(B_1 + B_2)^T)\chi & 0 \\ L_{31}(\chi_1) & L_{32}(\chi) & \chi_2 A_{c2} + A_{c2}^T \chi_2 \end{bmatrix} = \begin{bmatrix} r_{11} \\ r_{12} \\ r_{22} \end{bmatrix} \qquad (52)$$

Where

$$\bar{K}_1 = \begin{bmatrix} K_{11}^1 & K_{12}^1 \\ K_{12}^1 & K_{22}^1 \end{bmatrix} = \begin{bmatrix} \chi_1 & \chi \\ \chi^T & \chi_2 \end{bmatrix}$$

$$\begin{bmatrix} r_{11} & r_{12} \\ r_{12} & r_{22} \end{bmatrix} = K_0 \begin{bmatrix} B_2 \\ B_1 + B_2 \end{bmatrix} \{ \begin{bmatrix} B_2^T & (B_1+B_2)^T \end{bmatrix} K_0 \begin{bmatrix} 0 & I \\ 0 & I \end{bmatrix} + \begin{bmatrix} B_1^T & 0 \end{bmatrix} K_0 \begin{bmatrix} -I & 0 \\ 0 & -2I \end{bmatrix} \} + \{...\}^T \begin{bmatrix} B_2^T & (B_1+B_2)^T \end{bmatrix} K_0$$
$$+ K_0 \begin{bmatrix} B_1 B_2^T + B_2 B_1^T & B_1(B_1+B_2)^T \\ (B_1+B_2)B_1^T & 0 \end{bmatrix} K_0 \tag{53}$$

It holds:

$$r_{11} = \begin{bmatrix} I & 0 \end{bmatrix} \{ K_0 \begin{bmatrix} B_2 \\ B_1+B_2 \end{bmatrix} \{ \begin{bmatrix} B_2^T & (B_1+B_2)^T \end{bmatrix} K_0 \begin{bmatrix} 0 & I \\ 0 & I \end{bmatrix} + \begin{bmatrix} B_1^T & 0 \end{bmatrix} K_0 \begin{bmatrix} -I & 0 \\ 0 & -2I \end{bmatrix} \} + \{...\}^T \begin{bmatrix} B_2^T & (B_1+B_2)^T \end{bmatrix} K_0$$
$$+ K_0 \begin{bmatrix} B_1 B_2^T + B_2 B_1^T & B_1(B_1+B_2)^T \\ (B_1+B_2)B_1^T & 0 \end{bmatrix} K_0 \} \begin{bmatrix} I \\ 0 \end{bmatrix} = 0$$

This is easy to prove by just carrying out the multiplication using the right hand side of (53). Therefore to find $K_{11}^1$, one has to solve:

$$K_{11}^1 A_{c1} + A_{c1}^T K_{11}^1 = 0$$

and since we want an asymptotically stable closed loop system it will be that all the eigenvalues of $A_{c1}$ are negative, and thus $K_{11}^1 = 0$. Therefore the $K_1$ matrix will necessarily have the form

$$\bar{K}_1 = \begin{bmatrix} 0 & K_{12}^1 \\ K_{12}^1 & K_{22}^1 \end{bmatrix}$$

which is indefinite, except if also $K_{12}^1 = 0$ which is not necessary. We see that as the number of players goes to infinity, the cost of each player does not behave monotonically, although the part of the cost that depends on his initial condition $x_1(0)$ remains constant up a first order of $w$, i.e.:

$$J_1(w) \approx \frac{1}{2} \begin{bmatrix} x_1(0) & z(0) \end{bmatrix}^T (K_0 + w \begin{bmatrix} 0 & K_{12}^1 \\ K_{12}^1 & K_{22}^1 \end{bmatrix} + O(w^2)) \begin{bmatrix} x_1(0) \\ z(0) \end{bmatrix} \tag{54}$$

This is not true for the cost of the concatenated player who sees as his state $z$.

The case where $K_{12}^1 = 0$ is worthy examining since it would imply that the cost $J_1(w)$ behaves monotonically in $w$ up to order two. This holds only for very special values of the parameters of the game, and if it were to hold, it would mean that the players have the cost:

$$J_1(w) \approx \frac{1}{2} \begin{bmatrix} x_1(0) & z(0) \end{bmatrix}^T K_0 \begin{bmatrix} x_1(0) \\ z(0) \end{bmatrix} + w \frac{1}{2} z^T(0) K_{22}^1 z(0) + O(w^2) \left\| \begin{bmatrix} x_1(0) \\ z(0) \end{bmatrix} \right\|^2$$

where the first order with respect to $w$ change of the cost depends only on the "average" state $z$

**Comment 6**

The $\varepsilon(M,x(0))$ of the $\varepsilon(M,x(0))$-Nash Equilibrium defined earlier needs to be further qualified. From the analysis for $K_0, \bar{K}_1$ we see that:

$$\varepsilon(M,x(0)) \approx \frac{1}{2} w \begin{bmatrix} x_1(0) \\ w \sum_1^M x_i(0) \end{bmatrix}^T \bar{K}_1 \begin{bmatrix} x_1(0) \\ w \sum_1^M x_i(0) \end{bmatrix}$$

This makes meaningful the notion of $\varepsilon(M,x(0))$ equilibrium, since for $M$ very large, or equivalently for $w = \frac{1}{M}$ very small, the approximate Nash equilibrium approaches the exact one. Notice that if instead of using in (15) $K_0$ and $w = 0$ for calculating $L_1, L_2$ we use the $K_0 + w\overline{K}_1$, we are going to have again an $\varepsilon(M,x(0))$-Nash equilibrium where we will have a better approximation of order $w^2$.

**Comment 7**
The analysis of the present paper can be easily extended to the case where instead of the costs (2) we have costs of the form:

$$J_i = \frac{1}{2}\int_0^\infty (\begin{bmatrix} x_i \\ z \end{bmatrix}^T \begin{bmatrix} Q & Q_{21}^T \\ Q_{21} & Q_{22} \end{bmatrix} \begin{bmatrix} x_i \\ z \end{bmatrix} + u_i^T u_i + \sum_{j \neq i} u_j^T S_1 u_j + 2u_i^T S_2 \sum_{j \neq i} u_j + 2 \sum_{j<k, j\neq i, k \neq i} u_j^T S_3 u_k) dt$$

where the appropriate positive (semi)definiteness assumptions are made. Notice that this form preserves the symmetry and thus the solutions are again of the form (3).

## 4. A SCALAR EXAMPLE

In the model (1)-(2) we use $A_1 = a_1, A_2 = a, B_1 = 1, B_2 = b, Q = q, Q_f = q_f$, all scalars. Each $x_i$ evolves as

$$\frac{dx_i}{dt} = a_1 x_i + \frac{a}{M}(x_1 + x_2 + ... + x_M) + u_i + \frac{b}{M}(u_1 + u_2 + ... + u_M)$$

The cost of the i-th player is given by:

$$J_i = \frac{1}{2}\int_0^\infty (qx_i^2 + u_i^2) dt$$

The Nash solution for each player i will be of the form:

$$u_i = l_1 x_i + l_2 z = -[1+bw, (1+bw)] \begin{bmatrix} k_1 & k \\ k & k_2 \end{bmatrix} \begin{bmatrix} x_i \\ z \end{bmatrix}, K = \begin{bmatrix} k_1 & k \\ k & k_2 \end{bmatrix}$$

Where $K$ is the solution of the matrix Ricatti equation $R(K, w) = 0$,

$$A_c(0) = \begin{bmatrix} a_1 & a \\ 0 & a_1 + a \end{bmatrix} - \begin{bmatrix} b & 0 \\ 1+b & 0 \end{bmatrix} K_0 \begin{bmatrix} 0 & 1 \\ 0 & 1 \end{bmatrix} - \begin{bmatrix} 1 & 0 \\ 0 & 0 \end{bmatrix} K_0 = \begin{bmatrix} a_1 - k_1 & a - b(k_1+k) - k \\ 0 & a_1 + a - (1+b)(k_1+k) \end{bmatrix} = \begin{bmatrix} \lambda_1 & \lambda \\ 0 & \lambda_2 \end{bmatrix}$$

$A_{c1} = \lambda_1, A_{c2} = \lambda_2, A_{c0} = \lambda$

The equation $R_0(K_0) = 0$ results in the system:

$$-k_1^2 + 2a_1 k_1 + q = 0$$
$$-(1+b)y^2 + (2a_1 + a)y + q = 0$$
$$2k[a - by] - k^2 + 2k_2[a_1 + a - (1+b)y] = 0 \quad (55)$$
$$y = k_1 + k$$

The operator $L(K_0, X)$ can be written as a matrix multiplying a vector:

$$\Lambda \begin{bmatrix} \chi_1 \\ \chi \\ \chi_2 \end{bmatrix} = \begin{bmatrix} 2\lambda_1 \chi_1 \\ 2\lambda \chi_1 + 2\lambda_2 \chi - (k + b(k_1 + k))(\chi_1 + \chi) \\ 2\lambda_2 \chi_2 + 2\lambda \chi - 2(k_2 + (k_2 + k))(\chi_1 + \chi) \end{bmatrix} = \begin{bmatrix} 2\lambda_1 & 0 & 0 \\ \lambda - (k + b(k_1 + k)) & a_1 - (1+b)y & 0 \\ -2(k_2 + (k_2 + k)) & 2\lambda - 2(k_2 + (k_2 + k)) & 2\lambda_2 \end{bmatrix} \begin{bmatrix} \chi_1 \\ \chi \\ \chi_2 \end{bmatrix}$$

The important quantities are:

$\lambda_1 = a_1 - k_1$

$\lambda_2 = a_1 + a - (1+b)y$

$\bar{\lambda} = a_1 - (1+b)y$

Solving (55) we have obviously only one acceptable root for $\lambda_1 < 0$ and two possible roots for $\lambda_2$:

$k_1 = a_1 + \sqrt{a_1^2 + q}$

$\lambda_1 = -\sqrt{a_1^2 + q}$

$$y = k_1 + k = \frac{2a_1 + a + \varepsilon\sqrt{(2a_1 + a)^2 + 4q(1+b)}}{2(1+b)}, \varepsilon = \pm 1 \tag{56a}$$

$$\lambda_2 = \frac{a - \varepsilon\sqrt{(2a_1 + a)^2 + 4q(1+b)}}{2}$$

$$\bar{\lambda} = \frac{-a - \varepsilon\sqrt{(2a_1 + a)^2 + 4q(1+b)}}{2} \tag{56b}$$

For having $\lambda_2$ real we need:

$\Delta = (2a_1 + a)^2 + 4q(1+b) \geq 0$

For the Proposition 1 to hold we need: $\lambda_1 \neq 0, \lambda_2 \neq 0, \lambda_2 + \bar{\lambda} \neq 0$.

For Proposition 2 to hold we need: $\lambda_1 < 0, \lambda_2 < 0, \lambda_2 + \bar{\lambda} \neq 0$. Let us focus on the conditions for Proposition 2

**Case I**. For the closed loop system to be asymptotically stable and $L(K_0, X)$ invertible we need:

$\lambda_1 < 0, \lambda_2 < 0, \lambda_2 + \bar{\lambda} \neq 0$

The condition $\lambda_2 + \bar{\lambda} \neq 0$ amounts to $\Delta = (2a_1 + a)^2 + 4q(1+b) > 0$. We can have two acceptable $\lambda_2 < 0$, if

$0 < \sqrt{\Delta} < -a$. We can have only one acceptable $\lambda_2 = \dfrac{a - \sqrt{(2a_1 + a)^2 + 4q(1+b)}}{2}$ if $a(a_1 + a) + q(1+b) > 0$

**Case II**. For the closed loop system to be asymptotically stable, $L(K_0, X)$ invertible and the Nash solution to be "Stable" we need:

$\lambda_1 < 0, \lambda_2 < 0, \lambda_2 + \bar{\lambda} < 0$

The condition $\lambda_2 + \bar{\lambda} < 0$ amounts to $\varepsilon = 1$ and $\Delta = (2a_1 + a)^2 + 4q(1+b) > 0$. Therefore if the solution

$$\lambda_2 = \frac{a - \sqrt{(2a_1 + a)^2 + 4q(1+b)}}{2}$$

is negative it corresponds to an asymptotically stable closed loop matrix and the corresponding Nash solution is "Stable". If the other solution for $\lambda_2 = \dfrac{a + \sqrt{(2a_1 + a)^2 + 4q(1+b)}}{2}$ is also negative, it corresponds to another Nash solution that results to an asymptotically stable closed loop matrix, but this Nash solution is not "Stable".

According to Comment 4, we examine the problem:
$$\min J = \frac{1}{2}\int_0^\infty (q_e z_e(t)^2 + 2s_e z_e(t)u_e(t) + u_e(t)^2)dt$$
subject to:
$$\frac{dz_e}{dt} = (a_1 + a)z_e + (1+b)u_e$$
where it must be $q_e - s_e^2 \geq 0$ and $1 + b \neq 0$. It has as solution:
$$u_e = -[(1+b)\rho + s_e]z_e$$
where $\rho$ is the positive solution of the Ricatti equation:
$$0 = 2\rho(a_1 + a - (1+b)s_e) + q_e - s_e^2 - \rho^2(1+b)^2$$
We want $q_e, s_e$ chosen so that: $(1+b)\rho + s_e = k_1 + k$ or
$$y = k_1 + k = \frac{a_1 + a + \sqrt{(a_1 + a - (1+b)s_e)^2 + (1+b)^2(q_e - s_e^2)}}{1+b}$$
Recall that $y = k_1 + k$ satisfies (56a).

Equating the two expressions for $(k_1 + k)$ we find:
$$\sqrt{(a_1 + a - (1+b)s_e)^2 + (1+b)^2(q_e - s_e^2)} = -a + \varepsilon\sqrt{(2a_1 + a)^2 + 4q(1+b)} = -2\lambda_2$$

Notice that the left hand side has minimum value zero (achieved for $s_e = \dfrac{a_1 + a}{1+b}, q_e - s_e^2$) and can achieve any positive value for appropriately chosen $q_e, s_e, q_e \geq s_e^2$. The right hand side is achieved if the root $\lambda_2 < 0$ which will happen for the roots that yield an asymptotically stable closed loop matrix i.e. always when we have a Nash equilibrium.

## 5. CONCLUSIONS

In the present paper the limiting behavior of a dynamic Nash game was studied with respect to the number of players going to infinity. Similar questions for the discrete time LQ Nash game and for the Stackelberg equilibrium can be considered. The stochastic version of the problem studied here can also be considered for the case where the strategy $u_i$ is linear in the estimate of the average (market) state $z$ and the estimate of the state $x_i$. Of interest would be here to examine in the spirit of Ref.5 under what conditions better measurements (for example less measurement noise) are beneficial for the finite time horizon cases as the horizon increases and or as well as when the number of players does.

## 6. ACKNOWLEDGEMENT

This research has been cofinanced by the European Union (European Social Fund ESF) and Greek national funds through the Operational Program "Education and Lifelong Learning" of the National Strategic Reference Framework (NSRF) - Research Funding Program: THALES. Investing in knowledge society through the European Social Fund and the program ARISTEIA, project name HEPHAISTOS.